# A Domain-Specific Language for Discrete Mathematics

Rohit Jha          Alfy Samuel          Ashmee Pawar          M. Kiruthika

Dept of Computer Engineering
Fr. C.R.I.T., Vashi
Navi Mumbai, India

## ABSTRACT
This paper discusses a Domain Specific Language (DSL) that has been developed to enable implementation of concepts of discrete mathematics. A library of data types and functions provides functionality which is frequently required by users. Covering the areas of Mathematical Logic, Set Theory, Functions, Graph Theory, Number Theory, Linear Algebra and Combinatorics, the language's syntax is close to the actual notation used in the specific fields.

## General Terms
Discrete Mathematics, Programming Languages

## Keywords
Domain-Specific Language, Glasgow Haskell Compiler, Haskell, Preprocessor

## 1. INTRODUCTION

### 1.1 Domain-Specific Languages
A programming language can be defined as a language that is used to execute instructions and algorithms on a machine. These instructions or algorithms are represented as programs and have the properties of reliability, robustness, usability, portability, maintainability and efficiency.

A Domain-Specific Language (DSL) is a programming language that is targeted towards representing problems and the solutions of a particular domain or area [1]. By contrast, a General Programming Language (GPL) is used for developing software in a variety of application domains. Examples of commonly used DSLs are HTML, CSS, Verilog, LaTeX, SQL, AutoCAD and YACC. On the other hand, languages such as C, Java, Perl, Python and Ruby are examples of GPLs.

### 1.2 Characteristics of DSLs
A pervasive characteristic of DSLs is that they have a central and well-defined domain, allowing users to focus on the jargon of the problem domain, while screening away the complex internal operations of a system [2]. Since DSLs are used for a specific problem domain, they tend to have a clear notation for it, using meaningful symbols that are easy to enter using a keyboard or mouse. This results in a smooth learning curve for domain experts, who may not be adept in core programming skills. DSLs also empower them to easily comprehend and specify logic of their applications, and also maintain them with changing requirements. Thus, the popularity of a well-designed DSL lies in its capability of improving users' productivity and communication among domain experts.

### 1.3 When to create DSLs
Creating a DSL is worthwhile when the language allows particular types of problems or solutions to be expressed more clearly than what existing languages would allow, and also when the type of problem in question reappears sufficiently often. Repetitive tasks to be performed are readily defined in DSLs with custom libraries whose scopes are restricted to the domain. Hence, these tasks need not be defined from scratch each time. This increases users' productivity since DSLs require lesser time for programming and maintenance, as compared to GPLs.

### 1.4 Classification of DSLs
A recognized method to classify DSLs is to broadly categorize them as either internal or external [1]. An internal DSL is one that uses the infrastructure of a base or host language to build domain specific semantics on top of it. Internal DSLs are usually implemented in the form of a library for the base language and are also called embedded DSLs. It is preferable to develop internal DSLs if domain-specific constructs need not be strictly obeyed, or if domain-specific transformations and optimizations are not required [3]. External DSLs are developed as entirely new stand-alone DSLs, i.e. independent of a base language. This involves implementing stages such as lexical analysis, parsing, interpretation, compilation and code generation [4]. Thus, external DSLs have their own syntax and semantics.

### 1.5 Phases of DSL development
DSL development generally involves the following phases [3]: Decision, Analysis, Design, Implementation and Deployment. The decision phase is one in which the reasons for DSL development are weighed, with consideration of long-term goals along with economic and maintenance factors. In the analysis phase of DSL development, the problem domain is identified and domain knowledge is gathered. This requires input from domain experts and/or the availability of documents or code from which domain knowledge can be obtained. In the design phase, it is determined how the DSL would be implemented - whether it would be an internal or an external DSL. Following the design phase is the implementation phase, in which a suitable implementation approach is chosen. The DSL could be developed in the following approaches - interpreted, compiled, preprocessed, embedded, or even a hybrid of these. The DSL could be deployed in the form of library packages for base languages, or as source code to be built by the user, or even as a setup script along with installation files. While developing our DSL, we have followed guidelines mentioned in [5].

### 1.6 DSL for Discrete Mathematics
The domain of the developed DSL is discrete mathematics. The DSL consists of a library of functions and data structures for the branches of Set theory, Graph theory, Mathematical logic, Number theory, Linear algebra, Combinatorics and Functions. The language is a Preprocessed DSL, with the Haskell programming language as the base language and Glasgow Haskell Compiler (GHC) as the compiler. The reason for selecting Haskell is that it is purely functional and hence has no side effects. Haskell also provides a modern type system which incorporates features like type classes and generalized algebraic data types, giving it an edge over other languages. Like all functional programming languages, Haskell's notation is suited for mathematical representations. Apart from aiding mathematicians and physicists, the developed DSL would be useful in studying and describing objects and problems in branches of computer science, such as algorithms, programming languages, cryptography, automated theorem proving, and software development.

The layout of this paper is as follows: Section 2 describes the design of our DSL, including the benefits of functional programming and Haskell. In Section 3, the features of various modules for discrete mathematics included in the library are explained. Section 4 contains description of the library's modules, sample results of a few functions from these modules and outputs of applications developed using the DSL. Towards the end, in Sections 5 and 6, the paper is concluded after giving a view of the future scope.

## 2. DESIGN
## 2.1 Functional Programming Paradigm
In contrast to the imperative programming style, found in languages such as C, Java, Python and Ruby, the functional programming paradigm treats computation as the evaluation of mathematical functions and avoids state and mutable data [6]. Imperative functions suffer from side-effects, i.e. they can change the internal state of a program because they lack referential transparency. This means that an expression can result in different values at different times, depending on the state of the executing program. In functional languages devoid of side-effects, any evaluation strategy can be used, giving freedom to reorder or combine evaluation of expressions in a program. Since data is considered to be immutable, repeated modifications or updates to a value in functional programming languages lead to generation of new values every time.

As functional programming languages typically define programs and subroutines as mathematical functions, they are an ideal choice for developing mathematical tools. The advantage of using functional languages lies in the notion that they allow users to think mathematically, rather than rely on the workings of the underlying machine [7]. Besides, in the functional programming paradigm, functions are first-class objects, i.e. they can be passed as arguments to other functions, be returned from other functions and be assigned to variables and data structures. Functional languages provide the concept of higher order functions, which are functions that take other functions as inputs or return other functions as results. Moreover, functional languages result in shorter program codes, thereby easing maintenance and leading to higher programmer productivity.

## 2.2 Haskell
Apart from having the advantages of being a purely functional programming language, the benefits of using Haskell as a base language are [8] [9]:

### 2.2.1 Lazy Evaluation
Lazy evaluation or call-by-need is an evaluation strategy in which the evaluation of an expression is delayed until its value is actually needed and in which repeated evaluations are avoided. As a result, there is an increase in performance due to the avoidance of needless calculations and error conditions while evaluating compound expressions. Another advantage of lazy evaluation is construction of potentially infinite data structures. For example, the Haskell statement `a = [1..]` defines an infinite-length list of natural numbers. This feature lends itself to the creation of infinite sets in the DSL.

### 2.2.2 Expressive Type System
In Haskell, manipulation of complex data structures is made convenient and expressive with the provision of creating and using algebraic data types and performing pattern matching. Strong compile-time type checking makes programs more reliable, while type inference frees the programmer from the need to manually declare types to the compiler.

### 2.2.3 Smart Garbage Collection
Because Haskell is a purely functional language, data is immutable and all iterations of a recursive computation create a new value. Hence, computations produce more memory garbage than conventional imperative languages. This is easily handled in the DSL as GHC is efficient at garbage collection.

### 2.2.4 Polymorphic Types and Functions
Haskell supports parametric polymorphism and ad-hoc polymorphism. Parametric polymorphism refers to when the type of a value contains one or more (unconstrained) type variables, so that the value may adopt any type that results from substituting those variables with concrete types. Ad-hoc polymorphism refers to when a value is able to adopt any one of several types because it, or a value it uses, has been given a separate definition for each of those types. Polymorphism is defined for functions as well. This means that functions can take the same number of arguments as those of different data types. For example, a permutation function can take as input a list of integers, floating point numbers, strings or any other data type and the same set of operations would be performed on the input, irrespective of the type.

### 2.2.5 List Comprehensions
List comprehensions in Haskell bear close resemblance to the notation used for Set definitions. For instance, to obtain a list of squares of positive integers, the Haskell code is `squares = [x^2 | x <- [1..]]`. This is similar to the Set theory notation: $squares = \{x2 \mid x \in \{1, 2, \ldots \infty\}\}$

### 2.2.6 Extensibility
Haskell was built keeping in mind the extensibility required for modern functional programming languages. This allows creation of user defined functions, types, modules, etc. for the DSL.

## 2.3 Proposed Design Pattern
The disadvantage of embedded Domain Specific Languages is that their syntax and semantics are same as that of the base language. Thus, in order to use the embedded DSL, the users must be familiar with programming in the base language. The syntax of our DSL is kept close to the notation followed for discrete mathematics by implementing it as a Preprocessed Domain Specific Language. Apart from a library of modules for various concepts of discrete mathematics, the DSL includes a syntactic preprocessor which translates programs written in the DSL into equivalent Haskell representations. It is advantageous to use this approach for development as it allows the new language to have its own syntax, one which need not vary much from that of Haskell's.

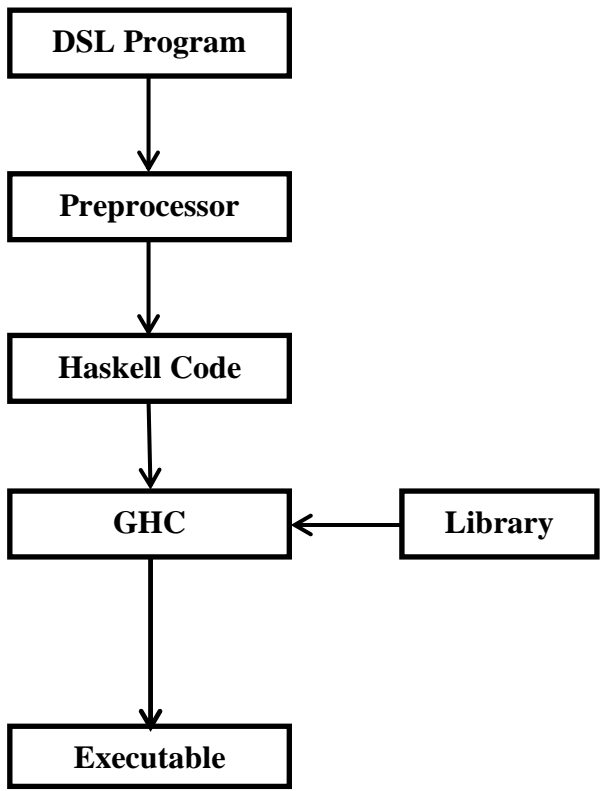

**Figure 1: Implementation design pattern for the DSL**

The proposed design pattern is shown in Figure 1. A preprocessor translates the source code written in the DSL into an equivalent Haskell representation. The generated Haskell code is then compiled using GHC, while importing required modules from the library, to produce an executable binary file. GHC conveniently provides command-line options for running a custom preprocessor over a source file [10]. If the preprocessor is named `cmd`, then compiling by using the options `-F -pgmF cmd`, followed by the source file's name, allows conversion of code in the source file into Haskell code. The preprocessor accepts at least three arguments: the first argument is the name of the original source file, the second is the name of the file holding the input, and the third is the name of the file where `cmd` should write its output to.

## 2.4 System Requirements
To be able to use the DSL comfortably, a user's system should be able to compile and run Haskell. For this, the system must have at least 128 MB of memory, 200 MB of disk space and GHC version 7.0.4 or later.

## 3. PROPOSED MODULES
## 3.1 Mathematical Logic
Logic is a vital topic of discrete mathematics, with applications in foundations of mathematics, formal logic systems and proofs. Often, set theory, model theory and recursion theory are considered as subsections of logic. In the DSL, logical operators and quantifiers from propositional logic, Boolean algebra and predicate logic are supported. This includes operators such as negation (NOT), conjunction (AND), disjunction (OR), exclusive disjunction (XOR), inverse conjunction (NAND), inverse disjunction (NOR), inverse exclusive disjunction (XNOR), logical implication (if...then), logical equality (iff), universal quantifier (for all) and existential quantifier (there exists some) and parentheses - '(' and ')'. Haskell provides a unary Boolean negation function (`not`) and binary operators for conjunction (`&&`) and disjunction (`||`), allowing development of other operators using these. Besides these, in Haskell, the universal and existential quantifiers are given by '`forall`' and '`exists`', respectively. The library module for mathematical logic also contains functions for applying the operations mentioned on lists of Boolean values.

## 3.2 Set Theory
According to Georg Cantor, the founder of set theory, a set is a gathering together into a whole of definite, distinct objects of our perception and of our thought - which are called elements of the set. The module currently focuses on naive set theory, operations on sets, relations, properties of relations and closures. Later, functionality would be added for groups, rings, fields, group-theoretic lattices and order-theoretic lattices, which find applications in cryptography and computational physics.

For sets, the module on set theory provides users with support for concepts such as checking for membership, empty/null set, subset, superset, generating power sets, finding cardinality, set difference, determining equality of sets, calculating Cartesian product, union of two sets, union of a list of sets, intersection of two sets, intersection of a list of sets, checking if two sets or a list of sets are disjoint, and mapping functions to sets. Working on sets is eased immensely with the provision of lists and list comprehension in Haskell.

Relations are sets of ordered pairs from elements of two sets, and are also called binary relations. The module for set theory in the library of the DSL contains functions for checking properties of relations. Important among these are those for checking if a relation is reflexive, symmetric, asymmetric, anti-symmetric, transitive, equivalence, partial order (weak or strict) and total order (weak or strict). With these as a base, functions for creating reflexive, symmetric and transitive closures are also developed and included in the library. As relations are essentially sets at their core, they can be combined by the operations of union, intersection, difference and composition. Composition also allows calculating powers of a relation and thus, the determination of transitive closures. The module also contains functions to check if a relation is a weak partial order, strong partial order, weak total order or strong total order.

## 3.3 Functions
Functions are algorithms or formulas which give certain values as output to parameters passed as input. The set of values that a function can take as input is called the domain of that function, and the set of values that a function can produce as its output is called the co-domain (or range) of that function.

Since the base language for the DSL is Haskell, a functional programming language, no additional support is required to be provided as such. To use composite functions in Haskell, a user has to simply represent them as (f . g) x, where f and g are the two functions, applied on the value x, and is read as "f of g of x". Composite functions can be made up of more than two functions. The constituent functions need only be separated by the dot operator. For example, (f . g . h . i) x is the composition of functions f, g, h and i.

## 3.4 Graph Theory

Considered the prime objects of study in discrete mathematics, and ubiquitous models for natural as well as man-made structures, graphs and trees are an important component of the DSL. This module provides support for users in computer science for studying networks, flow of computation, social network analysis, etc. In mathematics it would help users working with geometry, topology and group theory.

Graphs can be formally represented as the triple $G = (V, E, \phi)$, where V is a finite set of vertices, E is the finite set of edges and $\phi$ is the incidence function, with domain E and co-domain $P^2(V)$. Here, $P^2(V)$ represents the two-element subset of the power set P(V).

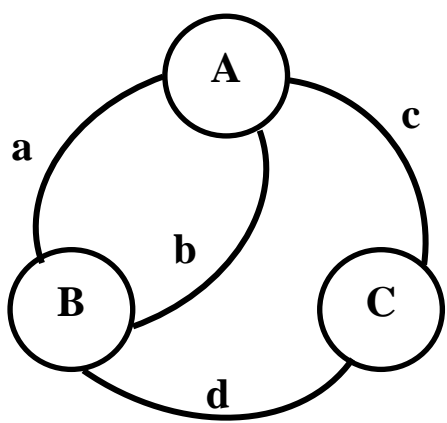

**Figure 2: An undirected graph with four nodes and three edges**

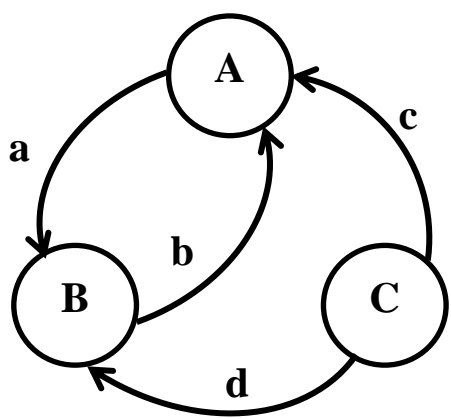

**Figure 3: A graph with directed edges**

The graph in Figure 2 is represented as $G = (V, E, \phi)$, such that V = {A, B, C}, E = {a, b, c, d} and $\phi$ = {(A, B), (A, B), (A, C), (B, C)}. Graphs may also be directed, in which case, the co-domain of $\phi$ would become V*V. An example of a directed graph can be $G = (V, E, \phi)$, such that V = {A, B, C}, E = {a, b, c, d} and $\phi$ = {(A, B), (B, A), (A, C), (C, B)}. This graph is shown in Figure 3.

Important graph operations such as finding in-degree and out-degree of vertices, finding nodes adjacent to a given node,

checking for cycles, calculating union of graphs, determining if a graph is a subgraph of another, finding existence of Euler paths, Euler circuits, Hamiltonian paths and Hamiltonian circuits are included in the module for graph theory in the DSL's library as well. Apart from these operations, the library also contains algorithms for Dijkstra's shortest path, Prim's and Kruskal's Minimum Spanning Tree algorithms, Depth-First Search, and Breadth-First Search.

This library module also contains functionality for Trees, primarily in the form of Binary Trees. It also contains frequently used functions such as in-order, pre-order and post-order tree traversals, inserting nodes in a tree, finding total number of nodes, searching for a particular node using Binary Search, determining height of a tree, checking if a tree is balanced and calculating depth of a node.

## 3.5 Number Theory

Number theory is one of the oldest and largest branches of mathematics. It primarily deals with the study of integers, but it also involves studying prime numbers, rational numbers and equations. Some applications of concepts in number theory are finding solutions to simultaneous linear equations, numerical analysis, group theory, field theory and elliptic curve cryptography.

The module for number theory covers generation of prime numbers using Sieve of Eratosthenes, primality testing using trial division and Miller-Rabin test, prime factorization of integers and random number generation. This module also contains functions for Fibonacci numbers, including generating a list of Fibonacci terms and finding the nth term of the Fibonacci series.

Elementary number theory consists of base/radix operations and manipulations. Accordingly, the module provides support for handling bases ranging up from 1 to any integer. This includes operations of addition, subtraction, multiplication, division and exponentiation in all bases, apart from conversion of numbers from a particular base to another.

Another important part of number theory is modular arithmetic. The DSL's library supports solving linear congruence relations of the form $ax \equiv b \pmod{m}$ and also evaluation of modular operations such as addition, multiplication and exponentiation.

## 3.6 Linear Algebra

The branch of linear algebra deals with vector spaces and linear mappings between these spaces. These are used to represent systems of linear equations in multiple unknowns. Combined with calculus, linear algebra facilitates the solution of differential equations. Linear algebra is applied in quantum mechanics, systems using the Fourier series, and several fields where simultaneous linear equations need to be solved.

The module for linear algebra in the DSL's library contains data structures for Vectors and Matrices, which are the essence of linear algebra. Vectors are represented as $n$-valued tuples `<v₁, v₂ ... vₙ>`, and $n \times m$ Matrices as `[row₁, row₂ ... rowₙ]`, where `rowᵢ = [aᵢ₁, aᵢ₂ ... aᵢₘ]` and $a_{ij}$ is an element. For example, consider the examples of a vector used in three-dimensional Cartesian system: `Vector <3,2,-7>` and the third order unit matrix: `Matrix [[1,0,0], [0,1,0], [0,0,1]]`. Operations such as finding the order of a matrix, calculating trace, transpose, determinant, inverse, multiplication, division, addition, subtraction and power of matrices are frequently applied in matrix theory, and functions for the same have been included in the library module.

The module also contains functions for checking properties of a matrix or whether a matrix is of a certain type. Some of these include checking if a matrix is symmetric, skew-symmetric, orthogonal, involutory, 0/1, unit/identity matrix, a zero matrix or a one matrix. In addition, a mapping function for matrices allows the application of a single function to all the elements of a matrix. The module contains functions for generating unit matrices of order $n$, $m{\times}n$ zero and one matrices.

For vectors, functions are developed for addition, subtraction, multiplication (scalar/dot/inner product), vector product, scalar triple product and vector triple product), calculating magnitude of a vector, calculating angle between two vectors, mapping a function to a vector, checking if a vector is a unit vector, determining order of vectors and extracting an element or even a range of elements from a vector. The module also contains functions to find sum and difference of a list of Vectors.

## 3.7 Combinatorics
This branch of mathematics deals with the study of countable discrete structures. This involves counting the structures, determining criteria, and constructing and analyzing objects satisfying these criteria. In computer science, combinatorics is used frequently in analysis of algorithms to obtain estimates and formulas.

For users involved in computational combinatorics, this DSL would be helpful as it has a module consisting of frequently used functions such as those to find factorials, permutations and combinations, generate permutation and combination lists and also to generate random permutations using the Fisher-Yates/Knuth shuffle algorithm.

## 4. IMPLEMENTATION AND RESULTS
This section describes modules from the DSL's library, including declaration of these modules and a few sample functions with results for every module. In addition, this section also contains results of applications developed using the DSL.

## 4.1 Mathematical Logic
The module for mathematical logic contains the following declaration for exporting functions to users' programs:

```
module MPL.Logic.Logic
(
        and',
        or',
        xor,
        xnor,
        nand,
        nor,
        equals,
        implies,
        (/\),
        (\/),
        (==>),
        (<=>),
        notL,
        andL,
        orL,
        xorL,
        xnorL,
```

```
        nandL,
        norL
)
where
```

Here, `MPL.Logic.Logic` is the module's name, indicating that the file is stored in the directory `MPL/Logic` and is named `Logic.hs`. This declaration is followed by definitions for each of the functions mentioned.

For example, consider the definition of the function for logical implication:

```
implies :: Bool -> Bool -> Bool
implies a b
        | (a == True)&&(b == False) = False
        | otherwise = True
```

In accordance with the objective of creating a notation close to the one actually used in discrete mathematics, an operator for logical implication is defined as follows:

```
(==>) :: Bool -> Bool -> Bool
a ==> b = implies a b
```

This provides syntactic sugar and improves readability. Now, the function for logical implication may be called by the user in any of the following three ways, all giving the same result - `False`:

```
implies True False
True `implies` False
True ==> False
```

As mentioned in 3.1, this module also defines functions which work on a list of Boolean values. The difference between the names of these functions and those of unary or binary functions is that they contain an additional 'L' as suffix, indicating that they operate on lists. A common operation is to find the XOR (Exclusive OR) of a list of values. Since the module already contains a function for finding the XOR of two values, it can be used to XOR the result of XOR of two values with the next value. Repeating this process for the length of the list gives a single final Boolean value. Such functions for lists of Boolean values are implemented using Haskell's `foldl1` function. The `xorL` function is defined as:

```
xorL :: [Bool] -> Bool
xorL a = foldl1 (xor) a
```

Here, `a` represents a list of `Bool`. An example of this function's usage is:

```
xorL [True, False, True, True, False]
```

This returns the `Bool` value `True`. The results of invoked functions and sample usage of operators are shown in Figure 12.

## 4.2 Set Theory
Under set theory, the library contains modules for working on Sets and Relations.

### 4.2.1 Sets
The module for Sets is declared as:

```
module MPL.SetTheory.Set
```

```
(
        Set(..),
        set2list,
        union, unionL,
        intersection,
        intersectionL,
        difference,
        isMemberOf,
        cardinality,
        isNullSet,
        isSubset,
        isSuperset,
        powerSet,
        cartProduct,
        disjoint,
        disjointL,
        sMap
)
where
```

The function `union` is defined as:

```
union :: Ord a => Set a -> Set a -> Set a
union (Set set1) (Set set2)
        = Set $ (sort . nub) (set1 ++ [e |
e <- set2, not (elem e set1)])
```

This is based on the definition that the union of two sets is the set containing all elements from that first set, and all elements from the second set that are not in the first. In addition, duplicates from this set are removed and this resultant set is sorted. If this function is called as `union (Set {2,4,6}) (Set {1,2,3})`, the output would be `Set {1,2,3,4,6}`.

A common set operation is that of finding the Cartesian product of two sets. In the library module, it is defined as:

```
cartProduct :: Ord a => Set a -> Set a ->
[(a,a)]
cartProduct (Set set1) (Set set2)
= Set [(x,y) | x <- set1', y <- set2']
        where
                set1' = (sort . nub) set1
                set2' = (sort . nub) set2
```

This function may be called as `cartProduct (Set {1,2}) (Set {3,4})` to produce the result `Set {(1,3),(1,4),(2,3),(2,4)}`.

In several conditions, it is requires to check if two sets are disjoint. For this, the module contains the function `disjoint`, and it is defined as:

```
disjoint :: Ord a => Set a -> Set a ->
Bool
disjoint (Set s1) (Set s2) = isNullSet $
intersection (Set s1) (Set s2)
```

If a user were to invoke this function as `disjoint (Set {1,3..10}) (Set {2,4..10})`, then he/she would get back `True` as the output. These results are shown in Figure 14.

### 4.2.2 *Relations*
The module for Relations is declared as:

```
module MPL.SetTheory.Relation
(
        Relation(..),
        relation2list,
        getFirst,
        getSecond,
        elemSet,
        returnFirstElems,
        returnSecondElems,
        isReflexive,
        isIrreflexive,
        isSymmetric,
        isAsymmetric,
        isAntiSymmetric,
        isTransitive,
        rUnion,
        rUnionL,
        rIntersection,
        rIntersectionL,
        rDifference,
        rComposite,
        rPower,
        reflClosure,
        symmClosure,
        tranClosure,
        isEquivalent,
        isWeakPartialOrder,
        isWeakTotalOrder,
        isStrictPartialOrder,
        isStrictTotalOrder
)
where
```

Consider the definition for the `isTransitive` function:

```
isTransitive :: Eq a => Relation a ->
Bool
isTransitive (Relation r)
= andL [(a,c) `elem` r | a <- elemSet r,
b <- elemSet r, c <- elemSet r, (a,b)
`elem` r, (b,c) `elem` r]
```

This function may be called by the user as `isTransitive (Relation {(1,1),(1,2),(2,1)})`, which would return `False`. However, the call `isTransitive (Relation {(1,1),(1,2),(2,1),(2,2)})` would return `True`. This result is shown in Figure 16.

The `symmClosure` function returns symmetric closure of the relation passed to it. It is defined as:

```
symmClosure :: Ord a => Relation a ->
Relation a
symmClosure (Relation r) = rUnion
(Relation r) (rPower (Relation r) (-1))
```

This function uses the property that symmetric closure of a relation is the union of that relation with its inverse. Calling the function as `symmClosure (Relation {(1,1),(1,3)})` would give the result as `Relation {(1,1),(1,3),(3,1)}`.

## 4.3  Graph Theory
Under graph theory, the library contains modules for Graphs and Trees.

### 4.3.1 Graphs

Declaration for the module on graphs is:

```
module MPL.GraphTheory.Graph
(
        Vertices(..),
        vertices2list,
        Edges(..),
        edges2list,
        Graph(..),
        GraphMatrix(..),
        graph2matrix,
        getVerticesG,
        getVerticesGM,
        numVerticesG,
        numVerticesGM,
        getEdgesG,
        getEdgesGM,
        numEdgesG,
        numEdgesGM,
        convertGM2G,
        convertG2GM,
        gTransposeG,
        gTransposeGM,
        isUndirectedG,
        isUndirectedGM,
        isDirectedG,
        isDirectedGM,
        unionG,
        unionGM,
        addVerticesG,
        addVerticesGM,
        verticesInEdges,
        addEdgesG,
        addEdgesGM,
        areConnectedGM,
        numPathsBetweenGM,
        adjacentNodesG,
        adjacentNodesGM,
        inDegreeG,
        inDegreeGM,
        outDegreeG,
        outDegreeGM,
        degreeG,
        degreeGM,
        hasEulerCircuitG,
        hasEulerCircuitGM,
        hasEulerPathG,
        hasEulerPathGM,
        hasHamiltonianCircuitG,
        hasHamiltonianCircuitGM,
        countOddDegreeV,
        countEvenDegreeV,
        hasEulerPathNotCircuitG,
        hasEulerPathNotCircuitGM,
        isSubgraphG,
        isSubgraphGM
)
where
```

As stated in section 3.4, the module contains functions which work on graphs defined both formally and as matrices. Functions for the former have 'G' as suffix, while functions for the latter have 'GM' as suffix. The implementation of functions for both is made possible by the functions `convertG2GM` and `convertGM2G`, which convert between the formal and matrix representations.

Consider the function for determining if a graph is undirected:

```
isUndirectedGM :: Ord a => GraphMatrix a
-> Bool
isUndirectedGM (GraphMatrix gm)
= (GraphMatrix gm) == gTransposeGM
(GraphMatrix gm)
```

When called as `isUndirectedGM (GraphMatrix [[0,5],[5,0]])`, `True` is returned. The invocation of functions for graphs is shown in Figure 17.

Using the property of that a graph has an Euler circuit only if all vertices have even degree, the function `hasEulerCircuitG` is defined as:

```
hasEulerCircuitG :: Ord a => Graph a ->
Bool
hasEulerCircuitG (Graph g)
= and [ even $ (degreeG (Graph g)
(Vertices [v])) | v <- vertices2list $
getVerticesG (Graph g)]
```

Thus, an invocation such as `hasEulerCircuitG (Graph (Vertices {1,2}, Edges {(1,2,4),(2,1,3)}))` would result in a return of `True`.

### 4.3.2 Trees

The module for trees has the following declaration:

```
module MPL.GraphTheory.Tree
(
        BinTree(..),
        inorder,
        preorder,
        postorder,
        singleton,
        treeInsert,
        treeSearch,
        reflect,
        height,
        depth,
        size,
        isBalanced
)
where
```

The functions `inorder`, `preorder` and `postorder` are functions for tree traversal. The definition for preorder is:

```
preorder :: BinTree a -> [a]
preorder Leaf = []
preorder (Node x t1 t2) = [x] ++ preorder
t1 ++ preorder t2
```

If we consider the following BinTree:

```
tree =
        Node 4
            (Node 2
                (Node 1 Leaf Leaf)
                (Node 3 Leaf Leaf))
            (Node 7
                (Node 5
                    Leaf
                    (Node 6 Leaf Leaf))
                (Node 8 Leaf Leaf))
```

Then the function call, `preorder tree`, would generate the result `[4,2,1,3,7,5,6,8]`.

In essence, the `BinTree` data type is a Binary Search Tree. The function `treeSearch`, is an implementation of the Binary Search algorithm and has the following definition:

```
treeElem :: Ord a => a -> BinTree a ->
Bool
treeElem x Leaf = False
treeElem x ( Node a left right )
      | x == a = True
      | x < a = treeElem x left
      | x > a = treeElem x right
```

The function `isBalanced` recursively checks if the height of all nodes at the same level are equal. The definition of this function makes use of the `height` function and is as follows:

```
isBalanced :: BinTree a -> Bool
isBalanced Leaf = True
isBalanced (Node x t1 t2) = isBalanced t1
&& isBalanced t2 && (height t1 == height
t2)
```

If this function is applied on `tree` as `isBalanced tree`, the output would be `False`. The results of functions for Trees are shown in Figure 13.

## 4.4 Number Theory
Under number theory, the library contains the following modules:

### 4.4.1 Base/Radix Manipulation
This module has the description:

```
module MPL.NumberTheory.Base
(
      toBase,
      fromBase,
      toAlphaDigits,
      fromAlphaDigits
)
where
```

The function `toBase` converts a decimal number into the equivalent form of a specified base/radix. It has the definition:

```
toBase :: Int -> Int -> [Int]
toBase base v = toBase' [] v where
      toBase' a 0 = a
      toBase' a v = toBase' (r:a) q where
(q,r) = v `divMod` base
```

When invoked as `toBase 8 37` or as `37 `toBase` 8`, the result would be `[4,5]`, which is read as 45, octal for 37. The result of `toBase` is also shown in Figure 15.

### 4.4.2 Fibonacci Series
The module on Fibonacci series contains two functions, `fib` and `fibSeries`. The function `fib` takes an integer as parameter and returns the term at that index in the Fibonacci series. It is defined as:

```
fib n = round $ phi ** fromIntegral n /
sq5
```

```
      where
            sq5 = sqrt 5 :: Double
            phi = (1 + sq5) / 2
```

If called as `fib 10`, the output is 55.

The `fibSeries` function takes an integer as parameter and returns the Fibonacci series as a list of integers. The definition is:

```
fibSeries n = [fib i | i <- [1..n]]
```

If a user wants to obtain the first 10 numbers in the Fibonacci series, he/she has to call the function as `fibSeries 10`, which gives the result `[1,1,2,3,5,8,13,21,34,55]`.

### 4.4.3 Modular Arithmetic
This module has the description:

```
module MPL.NumberTheory.Modular
(
      modAdd,
      modSub,
      modMult,
      modExp,
      isCongruent,
      findCongruentPair,
      findCongruentPair1
)
where
```

The `modExp` function is the function for modular exponentiation. It takes the numbers $a$, $b$ and $m$ as parameters and computes the value of $a^b$ mod $m$. The definition is:

```
modExp a b m = modexp' 1 a b
      where
      modexp' p _ 0 = p
      modexp' p x b =
            if even b
            then modexp' p (mod (x*x) m)
(div b 2)
            else modexp' (mod (p*x) m) x
(pred b)
```

If invoked as `modExp 112 34 546`, the integer `532` is returned.

### 4.4.4 Prime Numbers
This module has the following description:

```
module MPL.NumberTheory.Primes
(
      primesTo,
      primesBetween,
      firstNPrimes,
      isPrime,
      nextPrime,
      primeFactors
)
where
```

The function `primesTo` generates all prime numbers less than or equal to the number passed as parameter, using the Sieve of Eratosthenes. Its definition is:

```
primesTo :: Integer -> [Integer]
```

```
primesTo 0 = []
primesTo 1 = []
primesTo 2 = [2]
primesTo m = 2 : sieve [3,5..m]
```

The invocation `primesTo 20` gives the output: `[2,3,5,7,11,13,17,19]`.

## 4.5 Linear Algebra

Under linear algebra, the library has modules for Vectors and Matrices.

### 4.5.1 Vectors

This module's description is:

```
module MPL.LinearAlgebra.Vector
(
       Vector(..),
       vDim,
       vMag,
       vec2list,
       vAdd,
       vAddL,
       (<+>),
       vSub,
       vSubL,
       (<->),
       innerProd,
       (<.>),
       vAngle,
       scalarMult,
       (<*>),
       isNullVector,
       crossProd,
       (><),
       scalarTripleProd,
       vectorTripleProd,
       extract,
       extractRange,
       areOrthogonal,
       vMap,
       vNorm
)
where
```

The `vAngle` function returns the angle between two Vectors. It has the definition:

```
vAngle :: Floating a => Vector a ->
Vector a -> a
vAngle (Vector []) (Vector []) = 0
vAngle (Vector v1) (Vector v2) = acos (
(innerProd (Vector v1) (Vector v2)) / (
(vMag (Vector v1)) * (vMag (Vector v2)))
```

As shown in Figure 18, when invoked as `vAngle (Vector [1,1,1]) (Vector [2.5,2.5])`, the result is `0.6154797086703874`.

The function `scalarTripleProduct` is based on the functions `innerProduct` and `crossProduct`. It is defined as:

```
scalarTripleProd a b c = innerProd a
(crossProd b c)
```

To normalize a Vector, the `vNorm` function can be used. It has the definition:

```
vNorm (Vector v) = scalarMult (1/(vMag
(Vector v))) (Vector v)
```

If called as `vNorm (Vector [1,2,3])`, the output is the Vector: `<0.2672612419124244,0.5345224838248488,0.8017837257372732>`

### 4.5.2 Matrices

The module for matrices has the following description:

```
module MPL.LinearAlgebra.Matrix
(
       Matrix(..),
       mAdd,
       mAddL,
       (|+|),
       mSub,
       (|-|),
       mTranspose,
       mScalarMult,
       (|*|),
       mMult,
       mMultL,
       (|><|),
       numRows,
       numCols,
       mat2list,
       determinant,
       inverse,
       mDiv,
       (|/|),
       extractRow,
       extractCol,
       extractRowRange,
       extractColRange,
       mPower,
       trace,
       isInvertible,
       isSymmetric,
       isSkewSymmetric,
       isRow,
       isColumn,
       isSquare,
       isOrthogonal,
       isInvolutory,
       isZeroOne,
       isZero,
       isOne,
       isUnit,
       zero,
       zero',
       one,
       one',
       unit,
       mMap
)
where
```

The `mMult` function performs multiplication of two matrices and returns the resultant matrix. Its definition is:

```
mMult :: Num a => Matrix a -> Matrix a ->
Matrix a
```

```
mMult (Matrix m1) (Matrix m2) = Matrix $
[ map (multRow r) m2t | r <- m1 ]
        where
          (Matrix m2t) = mTranspose (Matrix
m2)
          multRow r1 r2 = sum $ zipWith (*)
r1 r2
```

To add syntactic sugar, the module exports the operator `|><|` for multiplying two matrices. Thus, if a user wishes to multiply a Matrix `m1`, which is `Matrix [[1,0],[0,1]]` and a Matrix `m2`, which is `Matrix [[4.5,8],[(-10),6]]`, he/she can call either `mMult m1 m2` or `m1 |><| m2`, to get the output as `Matrix [[4.5,8.0],[(-10.0),6.0]]`. The usage and result is shown in Figure 19.

Another common operation is to find inverse of a matrix. In this module, the function `inverse` is defined using the functions `cofactorM` and `determinant` as:

```
inverse (Matrix m) = Matrix $ map (map (*
recip det)) $ mat2list $ cofactorM
(Matrix m)
        where
          det = determinant (Matrix m)
```

If called as `inverse (Matrix [[1,1],[1,(-1)]])`, the result is the matrix: `Matrix [[0.5,0.5],[0.5,(-0.5)]]`.

The module contains several functions to check for properties of a matrix. One of these is `isOrthogonal`, which is to check if a matrix is orthogonal. Using the functions `mTranspose` and `inverse` it is easily defined as:

```
isOrthogonal (Matrix m) = (mTranspose
(Matrix m) == inverse (Matrix m))
```

When it is used as `isOrthogonal (Matrix [[1,1],[1.2,(-1.5)]])`, the output is `False`.

## 4.6 Combinatorics
This module has the description:

```
module MPL.Combinatorics.Combinatorics
(
        factorial,
        c,
        p,
        permutation,
        shuffle,
        combination
)
where
```

The function definition for `factorial` is:

```
factorial :: Integer -> Integer
factorial n
```

```
        | (n == 0) = 1
        | (n > 0) = product [1..n]
        | (n < 0) = error "Usage -
factorial n, where 'n' is non-negative."
```

This function can return arbitrarily large integers since its return type is `Integer`. When `factorial 5` is called, `120` is returned as the result.

The `factorial` function acts as a base for other functions in the module. For example, the function `p` returns the number of possible permutations of *r* objects from a set of *n* given by $_nP_r$. It is defined as:

```
p :: Integer -> Integer -> Integer
p n r = div (factorial a) (factorial (a-
b))
        where
                a = max n r
                b = min n r
```

When this function is called as `p 10 5` or `10 `p` 5`, `30240` is the output. Usage of these functions is shown in Figure 20.

## 4.7 Applications
This subsection contains sample results from applications developed using the DSL.

### 4.7.1 Ciphers
Fig. 4 and Fig. 5 show the output of a program developed in the DSL for enciphering and deciphering of messages using Caesar cipher and Transposition cipher.

### 4.7.2 RSA Encryption and Decryption
Sample execution results of a program for implementing RSA encryption system using the DSL are shown in Fig. 6 and Fig. 7. This program was developed using the library modules on modular arithmetic, `MPL.NumberTheory.Modular`, and on prime numbers, `MPL.NumberTheory.Primes`.

**Figure 4: Enciphering using Caesar cipher**

**Figure 5: Deciphering using Transposition cipher**

**Figure 6: Encryption using RSA**

**Figure 7: Decryption using RSA**

**Figure 8: Generating Shared Key using Diffie-Hellman Key Exchange protocol**

**Figure 9: Solving simultaneous linear equations**

### 4.7.3  Diffie-Hellman Key Exchange

A sample output of a program developed using the DSL for Diffie-Hellman Key Exchange protocol is shown in Fig. 8. The two users must have a common primitive root and a prime number. The shared key and public keys are calculated using the users' private keys.

### 4.7.4  Simultaneous Linear Equations

Since the DSL's library provides support for Matrices, it is easy to develop a program for solving simultaneous linear equations. Fig. 9 shows an output for solving the two equations in two variables: $x + 2y = 4$ and $x + y = 1$. It is also possible to solve linear equations in $n$ variables using $n$ or more equations.

### 4.7.5  Mersenne Prime Numbers

The DSL's module on prime numbers under number theory allows for efficient determination of Mersenne prime numbers. Mersenne prime numbers are prime numbers of the form $2^q$ - 1, where $q$ is also a prime number. Fig. 10 shows the output as a list of powers $q$, between 2 and 1000, which result in Mersenne primes. Fig. 11 shows the output as a list all Mersenne prime numbers less than $2^{100}$.

**Figure 10: Powers of Mersenne prime numbers**

**Figure 11: Mersenne prime numbers having powers from 2 to 100**

## 5. SCREENSHOTS OF IMPLEMENTED LIBRARY MODULES

**Figure 12: Module on Mathematical Logic in GHCi**

**Figure 13: Module on Trees in GHCi**

**Figure 14: Module on Sets in GHCi**

```
 File   Edit   View   Search   Terminal   Help
Prelude MPL.NumberTheory.Base> toBase 8 37
[4,5]
Prelude MPL.NumberTheory.Base> toBase 60 1024
[17,4]
```

**Figure 15: Module on Base Manipulation in GHCi**

```
 File   Edit   View   Search   Terminal   Help
*MPL.SetTheory.Relation> isTransitive (Relation [(1,1),(1,2),(2,1)])
False
*MPL.SetTheory.Relation> isTransitive (Relation [(1,1),(1,2),(2,1),(2,2)])
True
*MPL.SetTheory.Relation> symmClosure (Relation [(1,1),(1,3)])
{(1,1),(1,3),(3,1)}
```

**Figure 16: Module on Relations in GHCi**

```
 File  Edit  View  Search  Terminal  Help
Prelude MPL.GraphTheory.Graph> isUndirectedGM (GraphMatrix [[0,5],[5,0]])
True
Prelude MPL.GraphTheory.Graph> hasEulerCircuitG (Graph (Vertices [1,2], Edges [(1,2,4),(2,1,3)]))
True
```

**Figure 17: Module on Graphs in GHCi**

```
 File   Edit   View   Search   Terminal   Help
*MPL.LinearAlgebra.Vector> vAngle (Vector [1,1,1]) (Vector [2.5,2.5])
0.6154797086703874
*MPL.LinearAlgebra.Vector> vNorm (Vector [1,2,3])
<0.2672612419124244,0.5345224838248488,0.8017837257372732>
```

**Figure 18: Module on Vectors in GHCi**

```
 File   Edit   View   Search   Terminal   Help
Prelude MPL.LinearAlgebra.Matrix> (Matrix [[1,0],[0,1]]) |><| (Matrix [[4.5,8],[(-10),6]])
4.5     8.0
-10.0   6.0
Prelude MPL.LinearAlgebra.Matrix> inverse (Matrix [[1,1],[1,(-1)]])
0.5     0.5
0.5     -0.5
Prelude MPL.LinearAlgebra.Matrix> isOrthogonal (Matrix [[1,1],[1.2,(-1.5)]])
False
```

**Figure 19: Module on Matrices in GHCi**

```
 File   Edit   View   Search   Terminal   Help
*MPL.Combinatorics.Combinatorics> factorial 20
2432902008176640000
*MPL.Combinatorics.Combinatorics> 10 `p` 5
30240
*MPL.Combinatorics.Combinatorics> permutation [1,2]
[[1,2],[2,1]]
*MPL.Combinatorics.Combinatorics> combination 2 [["Head"],["Tails"]]
[["Head","Head"],["Head","Tails"],["Tails","Head"],["Tails","Tails"]]
```

**Figure 20: Module on Combinatorics in GHCi**

# 6. FUTURE SCOPE

Since discrete mathematics is a vast area of study, it is not possible to include all topics in the library during the initial stages of development. In the future, modules for group theory, information theory, geometry, topology and theoretical computer science can be added. Additionally, the preprocessor can be constantly updated to handle new modules and new features in existing ones. Apart from this, based on feedback and suggestions from users, the syntax of this DSL can be improved to suit their needs.

# 7. CONCLUSION

Discrete mathematics plays a central role in the fields of modern cryptography, social networking, digital signal and image processing, computational physics, analysis of algorithms, languages and grammars. The language developed, owing to its syntax, helps computer scientists and mathematicians to work in an easier and more efficient manner as compared to that while using a General Purpose Language (GPL). This language would also be helpful to those learning and teaching discrete mathematics.

# 8. REFERENCES

[1] Fowler, M. 2010, "Domain-Specific Languages", Addison-Wesley Professional.

[2] Taha, W. 2008, "Domain Specific Languages", IEEE International Conference on Computer Engineering and Systems (ICESS).

[3] Mernik, M., Heering, J., and Sloane, A. M. 2005, "When and How to Develop Domain-Specific Languages", ACM Computing Surveys (CSUR).

[4] Ghosh, D. 2011, "DSLs in Action", Manning Publications Co.

[5] Karsai, G., Krahn, H., Pinkernell, C., Rumpe, B., Schindler, M., and Volkel, S. 2009, "Design Guidelines for Domain Specific Languages", Proceedings of DSM 2009.

[6] Hughes, J., "Why Functional Programming Matters", The Computer Journal, 1989.

[7] Goldberg, B., "Functional Programming Languages", ACM Computing Surveys, Vol. 28, No. 1, March 1996.

[8] Hudak, P. 1996. "Building domain-specific embedded languages", ACM Computing Surveys, December 1996.

[9] http://www.haskell.org/haskellwiki/Introduction. Introduction to Haskell, retrieved on October 22, 2012.

[10] http://www.haskell.org/ghc/docs/7.0.4/html/users_guide/index.html. The Glorious Glasgow Haskell Compilation System User's Guide, Version 7.0.4